\documentstyle{EuroPhys}
\input epsf
\def\etal{{\hbox{{\tenit\ et al.\/}\tenrm :\ }}}

\def\And{{\rm and\ }}

\def\stars{\bigskip\centerline{***}\medskip}

\newif\ifboo \boofalse

\def\Review#1{\boofalse{\it #1},}
\def\Name#1{{\sc #1},}
\def\Vol#1{\ifboo Vol. {\bf #1}\else{\bf #1}\fi}
\def\Year#1{\ifboo #1\else(#1)\fi}
\def\Book#1{\bootrue{\it #1},}
\def\Page#1{\ifboo {\rm p. #1}\else{\rm #1}\fi}
\def\k{\kappa}
\def\l{\lambda}
\def\d{\delta}
\def\tsty{\textstyle}
\def\br#1{\langle #1\rangle}
\def\b#1{{\bf #1}}
\def\D{\Delta}

\def\s{\sigma}
\def\t{\tau}

\begin{document}
\euro{}{}{}{}
\Date{22 May 1997}
\shorttitle{B. P. Lee \etal CHARGE OSCILLATIONS IN ETC.}
\title{Charge oscillations in Debye-H\"uckel theory}
\author{Benjamin P. Lee\inst{1} \And Michael E. Fisher\inst{2} }
\institute{
	\inst{1} Polymers Division, National Institute of Standards and
Technology, Gaitherburg, Maryland 20899, USA\\
	\inst{2} Institute for Physical Science and Technology, University 
of Maryland, College Park, Maryland 20742, USA \\
}
\rec{}{}
\pacs{
\Pacs{61}{20.Qg}{Structure of associated liquids: electrolytes, molten 
salts, etc}
\Pacs{05}{20.$-$y}{Statistical mechanics}
\Pacs{05}{70.Jk}{Critical point phenomena}
}
\maketitle
\begin{abstract}
The recent generalized Debye-H\"uckel (GDH) theory is applied to the
calculation of the {\it charge-charge\/} correlation function, $G_{ZZ}(\b r)$.
The resulting expression satisfies both (i)  the charge neutrality condition
{\it and\/} (ii) the Stillinger-Lovett  second-moment condition for all 
$T$ and $\rho_N$, the overall ion density, and (iii) exhibits charge 
oscillations for densities above
a ``Kirkwood line'' in the $(\rho_N,T)$ plane.  This corrects the 
normally assumed DH charge correlations, and, when combined   
with the GDH analysis of the density correlations, leaves the GDH theory   
as the {\it only\/} complete description of ionic correlation functions, 
as judged by (i)--(iii), (iv) exact low-density $(\rho_N,T)$ variation, and
(v) reasonable behavior near criticality.
\end{abstract}

A complete theory of ionic fluids, and in particular one which describes
the ionic critical region, remains an outstanding challenge for statistical 
physics \cite{reviews}.
Recent progress has been made at the mean-field level by studies based on the
Debye-H\"uckel (DH) theory of 
the restricted primitive model (RPM) of
electrolytes \cite{DH}, supplemented 
with ion pairing, free-ion depletion, and the concomitant dipolar-ionic 
interactions \cite{FL}.  
A satisfactory theory must also include a description of the ionic
correlation functions.  However, the 
conventionally {\it assumed\/} DH ion-ion correlation
functions, of which there appear to be three varieties, have several 
shortcomings: the
most significant is the absence of a diverging density-density
correlation length at the DH critical point, which renders the 
theory totally uninformative as regards the relevant order-parameter 
fluctuations.  Further, although the predicted
charge-charge correlation function, $G_{ZZ}(\b r)$, with
the familiar screening decay as $e^{-\k_D r}/r$, is exact in the low-density 
limit, it violates the Stillinger-Lovett second-moment condition
\cite{SL}. 
%(claims to the contrary, e.g. \cite{McQuarrie}, are based implicitly
%on the low-density limit).
 Furthermore, there is much evidence indicating 
that the charge correlations become oscillatory at moderately 
high densities \cite{Kirkwood,Outhwaite}, a phenomenon also missed 
by the simple screening form.

Similar difficulties arise with the oft-used mean-spherical approximation 
(MSA), which exhibits, in particular, a complete cancellation of
Coulombic effects from the density-density correlation function,
$G_{NN}(\b r)$.  In this case some improvement has been found by 
adding an {\it ad hoc\/} term
%---with parameters chosen to ensure satisfaction  of certain sum rules---
to the assumed direct correlation functions and adjusting it to gain
consistency with various sum rules;
this generalized MSA, or GMSA, yields non-trivial density correlations,
including a diverging correlation length at the
MSA critical point \cite{Evans}.  

However, it fails badly at low densities \cite{LeeF,comment3,long} and, as 
explained elsewhere \cite{FL,LeeF,ZFL}, the MSA (and GMSA) appears
to be inferior to DH-based theories for describing the critical region.
Hence, we have sought to remedy the deficiencies of the DH correlation
functions as well, and specifically to do so in a more natural
way. 
By following the spirit of the DH approximation, we extended the
theory to the case of  non-uniform, slowly varying ionic densities, 
$\rho_\pm(\b r)$, thus
enabling derivation of a Helmholtz free energy {\it functional\/} 
\cite{LeeF,long}. 
Ion correlations may then be obtained by functional differentiation 
techniques.
This generalized Debye-H\"uckel (GDH) theory was applied to the
calculation of $G_{NN}(\b r)$, and provided not only the expected
critical divergence of the second-moment density correlation length, 
$\xi$, but also the surprising, {\it universal\/} and {\it exact\/} divergence 
of $\xi$ in the low-density limit \cite{LeeF,BF} (where the GMSA
fails by predicting a finite, non-universal value \cite{comment3}).  This
testament to the physical validity of the
GDH approach motivated the calculation of the
charge-charge correlations via GDH theory that is reported here.

We find an expression for $G_{ZZ}(\b r)\equiv
\br{\,[\rho_+(\b r)-\rho_-(\b r)]\,[\rho_+(\b 0)-\rho_-(\b 0)]\,}$ 
which in the low-density 
limit approaches the conventional and exact DH result, but which {\it also}
explicitly satisfies the Stillinger-Lovett second-moment condition.
Furthermore, it exhibits charge oscillations for densities above
a ``Kirkwood line'' in the density-temperature plane \cite{Evans}.
More concretely, we find for the RPM 
(equisized hard-spheres with diameters $a$, charges
$\pm q_0$, and solvent dielectric constant $D$) the closed-form, Fourier
transform expression for the charge-charge correlation function
\begin{equation}\label{Gqq}
\hat G_{ZZ}(\b k;\rho_N,T)=\rho_Nk^2/[\k_D^2+k^2 + a^{-2} g_0(\k_D a,k a)],
\end{equation}
where $\rho_N=\rho_++\rho_-$ is the total ion density while the Debye 
parameter is given, as usual, by $\k_D^2=4\pi q_0^2\rho_N/Dk_BT$,
and
\begin{equation}
g_0(x,q)= x^2(\cos q - 1) - [2\ln(1+x)-2x+x^2](\cos q-\sin q/q).
\end{equation}
Expansion in powers of $k$ yields
 %\begin{equation}
 %\hat G_{ZZ}(\b k;\rho_N,T)= {Dk_BT\over 4\pi q_0^2} k^2+O(k^4)
 %\end{equation}
$\hat G_{ZZ}(\b k;\rho_N,T)=(Dk_BT/4\pi q_0^2)\, k^2+O(k^4)$,
which demonstrates satisfaction of both the Stillinger-Lovett second-moment 
condition,
\begin{equation}\label{secondmoment}
\int d\b r\,r^2\, G_{ZZ}(\b r)=-6\rho_N/\k_D^2,
\end{equation}
and the ``zeroth-moment'' or charge-neutrality condition,
\begin{equation}\label{zerothmoment}
\int d\b r\, G_{ZZ}(\b r)=0\quad \Rightarrow\quad 
\int_{|\b r|>a} d\b r\, G_{ZZ}(\b r)= -\rho_N,
\end{equation}
for all $\rho_N$ and $T$.  In the low-density limit (\ref{Gqq})
becomes $\hat G_{ZZ}(\b k)\approx \rho_Nk^2/(\k_D^2+k^2)$, the exact, 
universal limiting behavior.  By analyzing the poles of (\ref{Gqq})
we may obtain the predicted large-distance behavior of 
$G_{ZZ}(\b r;\rho_N,T)$: from that we find that simple exponential decay 
persists only up to $x\equiv\k_D a= x_K$; for $x>x_K$ the decay is 
oscillatory.  Numerically we obtain the ``Kirkwood value''
\begin{equation}\label{xK}
x_K\simeq 1.17832,
\end{equation}
which lies in the usually expected range \cite{SL,Kirkwood,Outhwaite,Evans}.

Before presenting the GDH calculation,
we summarize briefly the conventional DH correlation functions for
comparison.  Debye-H\"uckel theory provides an approximate result
for $\phi_\s(\b r)$, the mean electrostatic potential at $\b r$ due to 
both a charge of type $\s$ fixed at the origin
and the corresponding induced charge distribution \cite{DH}, namely,
\begin{eqnarray}\label{DHphi}
\phi^{DH}_\s(\b r)&=& q_\s/D r - q_\s\k_D/D(1+\k_D a),\qquad r<a,\nonumber\\
&=& q_\s e^{\k_D(r-a)}/D(1+\k_D a),\qquad\quad\>\>\> r\geq a.
\end{eqnarray}
The DH thermodynamics uses only the self-potential 
%
%\begin{equation}
%\psi_\s(\b r')\equiv
$\lim_{\b r\to 0}[\phi_\s^{DH}(\b r)-q_\s/Dr]=q_\s\k_D/D(1+\k_Da)$.
%\end{equation}
In a first effort to obtain correlations, one may supplement the 
treatment of Debye and H\"uckel with the
formally exact statistical Poisson's equation, which for the RPM states
[2(b)]
\begin{equation}\label{statpoisson}
\nabla^2\phi_\s(\b r)=-(4\pi q_\s/D) G_{ZZ}(\b r)/\rho_N,\qquad(\s=\pm).
\end{equation}
This yields a $G^{\rm Poiss}_{ZZ}$ with a simple screening
decay that (i) satisfies (\ref{zerothmoment})
for all $\rho_N$ and $T$, but (ii) violates the
second-moment condition everywhere except in the low density, $\k_D a\to 0$
limit.  Furthermore, this Poisson route says nothing whatsoever about the
density-density correlations.

A second approach is to  parallel the derivation of (\ref{DHphi}) \cite{DH} by
putting $\phi^{DH}_\s(\b r)$ into a Boltzmann form,
$g_{\s\t}(\b r\neq\b 0)\equiv\br{\rho_\s(\b r)\rho_\t(\b 0)}/\rho_\s\rho_\t
\simeq\exp[-\beta q_\s\phi_\t(\b r)]$ for $r>a$  
with $g_{\s\t}(\b r)=0$ for $r<a$. 
Then both  $G^{\rm Bltz}_{ZZ}$ and $G^{\rm Bltz}_{NN}$ may be 
obtained, although the latter displays no sign of the proper critical 
behavior.  However,  a more glaring defect of this
approach is that $G^{\rm Bltz}_{ZZ}(\b r)$ violates not only the 
second-moment condition, but {\it also\/} the charge-neutrality sum
rule! [This follows readily from the inequality $\sinh x > x$
(when $x>0$).]

A third path, perhaps the most travelled in the literature, is to linearize 
the Boltzmann form to get $g_{\s\t}(\b r)\simeq 1-\beta q_\s\phi_\t(\b r)$,
for $r>a$.  This gives
the same charge correlations as does the Poisson route, but for 
$G_{NN}^{\rm lin}$ the Coulombic terms cancel completely, as in the MSA.

Now, although (\ref{statpoisson}) is an exact relation,
approximate theories are generally inconsistent with respect to some
identities; indeed, only the exact solution can satisfy all possible
relations.  Our GDH theory \cite{LeeF} provides an alternate, formally 
exact approach to the correlations, which more closely follows the 
thermodynamic theory and may then be judged on its relative merits.

Our guiding motivation is that a free-energy-functional formulation
ensures density correlations that 
are sensitive to the critical point (i.e.,\ the compressibility relation
is satisfied by construction); thus, we are led to develop a DH 
theory for an inhomogeneous electrolyte.  As detailed elsewhere 
\cite{long}, the approach is sufficiently
general to allow the calculation, via functional differentiation 
techniques, of all the $\s,\t$ correlation functions $G_{\s\t}(\b r)$ for an 
arbitrary multi-component, equisized hard-sphere
electrolyte.  For brevity, however, we restrict consideration
here to the RPM, for which the {\it imposed\/} density variations
\begin{equation}\label{denvar}
\rho_\pm(\b r)=\bar\rho_\pm [1\pm\Delta\cos\b k\cdot\b r],
\end{equation}
(i.e.,\ in response to an external potential)
result in the Helmholtz free energy 
\begin{equation}\label{delF}
\bar f[\rho_\s(\b r)]-\bar f[\bar\rho_\s]
=-{\tsty {1\over 4}}\rho_N^2\hat G_{ZZ}^{-1}(\b k)\D^2+O(\D^4),
\end{equation}
where $\bar f\equiv -\beta F/V$ \cite{LeeF,long}.  Thus $G_{ZZ}$ may be 
found by expansion in $\D$.

The generalized Debye charging process yields the free energy
 \cite{long,comment2}
\begin{equation}\label{charging}
F[\rho_\s(\b r)]=F^{HC}[\rho_\s(\b r)]+\int_0^1 d\lambda\sum_\s\int d\b r' 
\rho_\s(\b r')\psi_\s(\b r',\{\l q_\s\}).
\end{equation}
where $F^{HC}$ denotes the pure hard-core Helmholtz free
energy functional while the mean electrostatic potential seen by an
ion of type $\s$ fixed at $\b r'$ is found from
\begin{equation}
\psi_\s(\b r')\equiv\lim_{\b r\to\b r'}[\phi_\s(\b r;\b r')-
q_\s/D|\b r-\b r'|],
\end{equation}
in which $\phi_\s(\b r;\b r')$ is the potential 
 at $\b r$ due to {\it both\/} the fixed charge at $\b r'$ 
{\it and\/} the induced charge distribution \cite{LeeF,long}.
[Compare with (\ref{DHphi}) above, {\it et seq}\/.]
To calculate $\phi_\s(\b r;\b r')$ we begin with the
exact {\it inhomogeneous} version of the statistical Poisson's equation
\cite{long},
\begin{equation}\label{ispe}
\nabla_r^2\phi_\s(\b r;\b r')=-(4\pi/D)\sum_\t q_\t\rho_\t(\b r)g_{\t\s}(\b r;
\b r').
\end{equation}
The DH approximation is to replace the $g_{\t\s}$ with Boltzmann factors
which depend on the potential.  However, it is crucial to note that
the varying ionic charge densities carry along an imposed overall
electrostatic potential $\Phi(\b r)$, determined simply by
\begin{equation}\label{Phidef}
\nabla^2\Phi(\b r)=-4\pi q_0\rho_Z(\b r)/D\equiv 
-(4\pi/D)\sum_\s q_\t\rho_\t(\b r),
\end{equation}
with appropriate boundary conditions.  This is independent of
the fixed charge of type $\s$ at $\b r'$, and therefore should not 
contribute to the Boltzmann factor for $g_{\t\s}$.  Hence, in the spirit
of DH, we take
\begin{equation}\label{GDHboltzmann}
g_{\t\s}(\b r;\b r')\simeq\exp[-\beta q_\t\widetilde\phi_\s(\b r;\b r')], 
\qquad |\b r-\b r'|>a,
\end{equation}
with the ``local induced potential''
\begin{equation}\label{phisep}
\widetilde\phi_\s(\b r;\b r')\equiv\phi_\s(\b r;\b r')-\Phi(\b r),
\end{equation}
and, as before, $g_{\t\s}=0$ for $|\b r-\b r'|<a$.  The need for
the separation of $\phi_\s(\b r;\b r')$ into a background
$\Phi(\b r)$ and an induced piece $\widetilde\phi_\s(\b r;\b r')$ is clear
in the limit $|\b r-\b r'|\to\infty$,
in which $\ln[g_{\t\s}(\b r;\b r')]$ must vanish while
$\phi_\s(\b r;\b r')\to\Phi(\b r)$ \cite{comment1}.

Next one inserts (\ref{GDHboltzmann}) into (\ref{ispe}) and makes
the second approximation of the DH procedure, namely linearization.  
This results in the full GDH equation
\begin{eqnarray}\label{GDHeq}
\nabla^2_r\widetilde\phi_\s(\b r;\b r')&=&
-(4\pi/D)\,\bigl[q_\s\d(\b r-\b r')-q_0\rho_Z(\b r)\bigr], 
\quad |\b r-\b r'| \leq a,\nonumber\\
&=& \tilde\k_D^2(\b r)\widetilde\phi_\s(\b r;\b r'),\qquad\qquad\qquad 
\qquad\>\>\>\>\>  |\b r-\b r'| \geq a,
\end{eqnarray}
where the spatially varying Debye parameter \cite{Outhwaite,LeeF} is
defined by
$\tilde\k_D^2(\b r)\equiv(4\pi/D)\sum_\t q_\t^2\rho_\t(\b r)$. 
The second term on the righthand side for $|\b r-\b r'| \leq a$, i.e.,\ 
{\it inside\/} the hard sphere, is needed to cancel the background charge
density $\rho_Z(\b r)$ there; it represents an effective ``cavity source''
term.

To obtain $G_{ZZ}$ for the RPM we chose the spatially varying densities 
(\ref{denvar}), for
which $\tilde\k_D^2(\b r)$ reduces simply to $\k_D^2$. The resulting GDH
equation (\ref{GDHeq}) can be solved readily by
Green's function methods \cite{LeeF,long}.
Integrating the self-potential $\psi_\s(\b r')$ against the density and 
charging according to (\ref{charging})
gives the free energy to order $\D^2$, from which our
main result (\ref{Gqq}) follows by use of (\ref{delF}).  Note that the 
$q_\s\d(\b r-\b r')$ source does not contribute directly to the charge-charge
correlation function in the simple case of the RPM; rather it is the 
cavity term and $\Phi(\b r)$ that serve to determine $G_{ZZ}$.

To elucidate the long distance behavior of $G_{ZZ}(\b r)$ we solve for
the pole, $k_0$, of $\hat G_{ZZ}(\b k)$
that lies closest to the origin in the complex $k$ plane.  
The real and imaginary parts of $k_0$ plotted in fig.\ 1 were found by 
solving the coupled equations ${\rm Re}[\hat G_{ZZ}^{-1}
(k_0)]={\rm Im}[\hat G_{ZZ}^{-1}(k_0)]=0$ numerically, using the
Newton-Raphson method.  When this 
pole is purely imaginary, corresponding to the leftmost part of 
curve (a) in fig.\ 1, $G_{ZZ}$ decays monotonically as
$e^{-r/\xi_Z}/r$, where the screening length is 
$\xi_Z=1/{\rm Im}(k_0)$.    In the low-density limit one finds 
$\xi_Z^{-1}=\k_D[1+{\tsty{1\over 4}}x^2-\frac 1 9 x^3 + \frac{19}{96}x^4+
\dots]$ so that $\xi_Z$ 
approaches the Debye value $\xi_D\equiv 1/\k_D$: see curve (c).  As $\k_D$
and $\rho_N/T$ increase, the pole $k_0$ and a nearby, purely 
imaginary pole $k_1$, curve (d), approach:  at
the Kirkwood value, $\k_D a=x_K$ [see (\ref{xK})], they merge, 
with $a/\xi_Z=2.266_0$, and for 
larger $\k_D$ they become complex, implying the oscillatory decay
$G_{ZZ}(\b r)\sim\cos[(2\pi r/\l)-\theta_{\k_D}]e^{-r/\xi_Z}/r$, with
 $\lambda=2\pi/{\rm Re}(k_0)$ [see plot (b)] and $\theta_{\k_D}$ a phase 
shift.  Hence, charge oscillations occur for densities
$\rho_N^*=\rho_Na^3 > x_K^2 T^*/4\pi \simeq 0.110\, T^*$,
where $T^*\equiv Dk_BTa/q_0^2$.  Finally, at $x=x_X=6.6523_2$ the
poles move to the real axis, i.e.,\ they merge with their complex conjugates.
Here the oscillations, with wavelength $\lambda_X \simeq 1.847a$, are no 
longer damped; this is suggestive of the onset of
crystallization \cite{Outhwaite} although this region certainly lies beyond 
the limits of validity of our approximation.

%\end{multicols}

\begin{figure}
\centerline{
\epsfxsize=5.5truein
\epsfbox{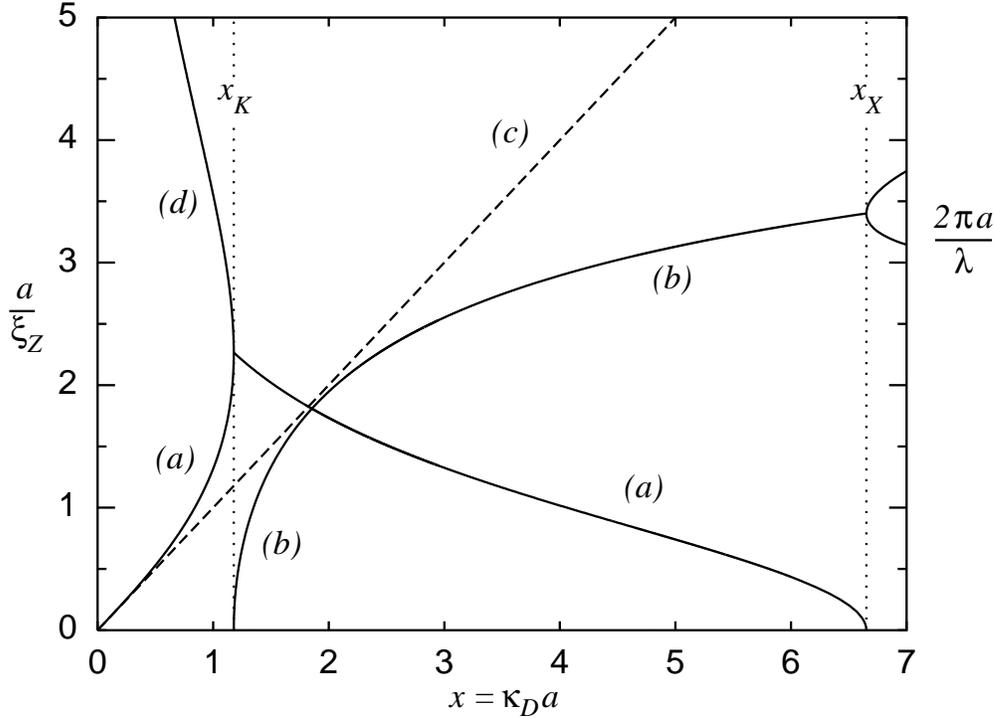}}
\medskip
\caption{Plots versus $x=(4\pi q_0^2\rho_Na^2/Dk_BT)^{1/2}$ of 
(a)~Im$(k_0a)=a/\xi_Z$, the inverse charge-charge correlation
length, (b)~Re$(k_0a)=2\pi a/\lambda$, the inverse charge oscillation 
wavelength, (c)~$x=\k_Da$, the inverse Debye length, and (d)~Im$(k_1a)$, 
the subleading pole.  Note that within pure DH theory the critical point 
occurs at $x_c=1$ and $T_c^*=1/16$; in pairing theories based on
DH theory $x_c=0.9$--$1.1$ and $T_c^*=0.052$--$0.057$ [1,3,15].
}
\label{fig1}
\end{figure}

As a consequence of our analysis one sees that
the GDH theory for ion correlations is the only one available that
satisfies the requisite 
sum rules (\ref{secondmoment}) and (\ref{zerothmoment}), gives
exact results for $G_{ZZ}(\b r)$ {\it and\/} $G_{NN}(\b r)$ in
the low density limit \cite{LeeF,BF}, and behaves sensibly in the
critical region, predicting (mean-field-like) diverging density
fluctuations \cite{LeeF}.

Finally, we remark on the addition of dipolar ion pairing.  Following 
\cite{FL} and \cite{LeeF}, one may straightforwardly 
add pairing to the calculation of $G_{ZZ}$.  However, in this case, as 
opposed to the density-density
correlations, one expects there to be no major contribution from
the pairs, since, in the center-of-mass approximation, the dipolar
ion pairs, of density $\rho_2$, appear as neutral objects that cannot 
contribute directly to the charge-charge correlations.
One important role of pairing, however, concerns the location of the
Kirkwood line in the $(\rho_N,T)$ plane: owing to the {\it depletion\/} 
of the free ion density, $\rho_1\equiv \rho_++\rho_- = \rho_N-2\rho_2$, 
the onset of charge oscillations
is expected to lie close to the critical region: see \cite{ZFL,GC} for 
the critical parameters and degree of pairing in the various approximations.
Lastly, we mention that the charge-charge correlations in the GDH formulation
with pairing still satisfy the sum rules (\ref{secondmoment}) and 
(\ref{zerothmoment}); however, in the Stillinger-Lovett rule 
(\ref{secondmoment}) the ``background'' dielectric constant $D$ that enters
the definition of $\kappa_D$ acquires a linear dependence on 
$\rho_2(\rho_N,T)$.  In general, the state-dependence of $D$ seems an open
question \cite{SL} although most authors seem to hold that it should be
totally absent in the RPM.  If that is correct the pairing treatment would
require further improvement.  (See also \cite{ZFL}.)

\stars

The insight underlying this work arose from discussions with
Professor Jean Pierre Hansen and from his lectures and
lecture notes which are gratefully acknowledged.  The interest of
Dr.\ Jack Douglas, Dr.\ Stefan Bekiranov and Daniel M. Zuckerman has
been appreciated.  Our researches have been 
supported by NSF Grant Nos. CHE 93-11729 and 96-14495 and by a
National Research Council Research Associateship.


\begin{thebibliography}{99}
% 
\bibitem{reviews} See, e.g.,
\Name{Fisher M.E.} \Review{J. Stat. Phys.} \Vol{75} \Year{1994} \Page{1};
{\it ibid.} \Review{J. Phys. Condens. Matt.} \Vol{8} \Year{1996} \Page{1}.
% 
\bibitem{DH}
(a) \Name{Debye P. \And H\"uckel E.}
\Review{Phys. Z.} \Vol{24} \Year{1923} \Page{185}. For
a modern presentation, and the low-density results, see
(b) \Name{D. A. McQuarrie} \Book{Statistical Mechanics} 
(Harper-Collins, NY, 1976), chap.\ 15.
% 
\bibitem{FL}(a) 
\Name{Fisher M.E. \And Levin Y.} \Review{Phys. Rev. Lett.} \Vol{71} 
\Year{1993} \Page{3826}, (b) \Name{Levin Y. \And Fisher M.E.} 
\Review{Physica A} \Vol{225} \Year{1996} \Page{164}.
% 
\bibitem{SL}(a) 
\Name{Stillinger F.H. \And Lovett R.} \Review{J. Chem. Phys.} \Vol{48}
\Year{1968} \Page{3858}, (b) {\it ibid.\/} \Review{J. Chem. Phys.} \Vol{49}
\Year{1968} \Page{1991}.
% 
\bibitem{Kirkwood}
\Name{Kirkwood J.G.} \Review{Chem. Rev.} \Vol{19} \Year{1936} \Page{275},
found $x_K \simeq 1.03$.
% 
\bibitem{Outhwaite}
\Name{Outhwaite C.W.} in \Book{Statistical Mechanics, A Specialist
Periodical Report} edited by \Name{K. Singer} \Vol{2}
(London: The Chemical Society) p\Page{188}; the linearized modified
Poisson-Boltzmann (MPB) theory yields $x_K \simeq 1.241$.
% 
\bibitem{Evans}
\Name{Leote de Carvalho R.J.F. \And Evans R.} (a) \Review{Molec. Phys.} 
\Vol{83} \Year{1994} \Page{619}, (b) \Review{J. Phys. Condens. Matter}
\Vol{7} \Year{1995} \Page{L575}; analysis of the GMSA gives $x_K \simeq
1.228$.
% 
\bibitem{LeeF}
\Name{Lee B.P. \And Fisher M.E.} \Review{Phys. Rev. Lett.} \Vol{76} 
\Year{1996} \Page{2906}.
% 
\bibitem{comment3}
In \cite{LeeF} replace the first part of equation (13) by $\xi_{\rm
GMSA}\approx({1\over 8}ab)^{1/2}$ and modify the following sentence by
replacing $\xi_{\rm GMSA}$ with $\xi_{\infty\rm GMSA}$.
%
\bibitem{long}
\Name{Lee B.P. \And Fisher M.E.} [to be published].
% 
\bibitem{ZFL}
\Name{Zuckerman D.M., Fisher M.E. \And Lee B.P.} [submitted for publication].
%
\bibitem{BF}
\Name{Bekiranov S. \And Fisher M.E.} [in preparation] based on the analysis
of \Name{Meeron E.} \Review{J. Chem. Phys.} \Vol{28} \Year{1958} \Page{630}.
% 
\bibitem{comment2}Note that all but the ideal gas 
components of $F^{HC}$ cancel out of (\ref{delF}). 
%
\bibitem{comment1} For locally charge-neutral 
density variations, as considered in \cite{LeeF}, $\Phi(\b r)$ vanishes
and the separation (\ref{phisep}) is unnecessary.
%
\bibitem{GC}
\Name{Fisher M.E. \And Lee B.P.} \Review{Phys. Rev. Lett.} \Vol{77}
\Year{1996} \Page{3561}.
%
\end{thebibliography}
\end{document}